# Short-time Fourier transform based on stimulated Brillouin scattering

Pengcheng Zuo[a,b], Dong Ma[a,b], and Yang Chen[a,b,*]

[a] *Shanghai Key Laboratory of Multidimensional Information Processing, East China Normal University, Shanghai 200241, China*
[b] *Engineering Center of SHMEC for Space Information and GNSS, East China Normal University, Shanghai 200241, China*
[*] *ychen@ce.ecnu.edu.cn*

**ABSTRACT**

In this paper, all-optical short-time Fourier transform (STFT) based on stimulated Brillouin scattering (SBS) is proposed and further used for real-time time-frequency analysis of different radio frequency (RF) signals. In the proposed all-optical STFT scheme, SBS not only provides a band-pass filter for implementing the window function in conjunction with a periodic frequency-sweep optical signal but also obtains the frequency domain information in different time windows through the generated waveform via frequency-to-time mapping (FTTM). A periodic frequency-sweep optical signal is generated and then modulated at a Mach-Zehnder modulator by the electrical signal under test (SUT). During different sweep periods, the fixed Brillouin gain functions as a bandpass filter to select a specific range of the spectrum, which is equivalent to applying a sliding window function to the corresponding section of the temporal signal with the help of the sweep optical signal. At the same time, after the optical signal is selectively amplified by the SBS gain and converted back to the electrical domain, SBS also implements the real-time FTTM, which can be utilized to obtain the frequency domain information corresponding to different time windows through the generated waveforms via the FTTM. The frequency domain information corresponding to different time windows is formed and spliced to analyze the time-frequency relationship of the SUT in real-time. An experiment is performed. STFTs of a variety of RF signals are carried out in a 12-GHz bandwidth limited only by the equipment, and the dynamic frequency resolution is better than 60 MHz.

## 1. Introduction

Characterization of unknown microwave signals plays a significantly important role in the entire microwave system and microwave system-based applications, such as high-speed communication systems, radar systems, and electronic warfare systems. In general, instantaneous frequency measurement (IFM) is fundamentally required to recognize single- or multi-frequency microwave signals, whereas spectrum analysis and the acquisition of time-frequency information are always required to analyze broadband signals. Among them, one of the typical time-frequency analysis methods, named short-time Fourier transform (STFT), is widely utilized to obtain the time-frequency information of microwave signals, especially for non-stationary signals [1], [2]. The key of STFT is local stationarity, in which the long non-stationary signal is regarded as the superposition of a series of short-term stationary signals by adding a finite sliding time window function [3], [4]. Thanks to the unique observation perspective offered by STFT, it has attracted great attention in many fields [5]-[16], especially widely used in the fields of speech, image processing, and radar signal processing. At this stage, STFT in these works is conventionally accomplished in the electrical domain. However, limited by the sampling rate of the analog-to-digital converter (ADC), the processing speed of STFT based on digital signal processing (DSP) in the electrical domain is relatively slow, resulting in that the analysis bandwidth and frequency are inherently limited [17], [18].

During the past two decades, microwave photonics (MWP) has focused on the generation, processing, control, and measurement of microwave signals and a number of promising solutions have been provided by it, taking the advantage of large bandwidth, high frequency, good tunability and immunity to electromagnetic interference offered by modern optics [19]-[26]. It has been proved that photonics-assisted spectral measurement [27], [28] and Fourier analysis [29] for microwave signals have better performance in terms of analysis frequency range and bandwidth compared with the electrical-based solutions. To date, however, only a few dispersion-based attempts for implementing the all-optical STFT have been proposed to overcome the bottleneck of the STFT in the electrical domain [30]-[32]. In [30], an all-optical STFT approach based on a temporal pulse shaping system incorporating an array of cascaded linearly chirped fiber Bragg gratings (LCFBGs) was proposed. STFT of an electrical signal consisting of two cascaded signals was experimentally demonstrated. However, the frequency resolution is limited by the dispersion value. Moreover, the number of frequencies that can be analyzed is related to the number of LCFBGs, which severely limits its practicality. In [31], a simpler proposal for implementing STFT was performed by using a short-pulse sampling of the signal under test (SUT) followed by linear dispersive propagation. However, the 340-MHz frequency resolution demonstrated in [31] still depends on a large dispersion value of around 6825 $ps^2$ $rad^{-1}$. This is equivalent to using a dispersion compensation fiber (DCF) of approximately 50 km, which may not be conducive to the real-time nature of the system. In addition, the implementation of STFT needs to meet the condition that the relative time delay between consecutive spectral copies of the SUT must be equal to the sampling period, which is not conducive to reconfigurability. The STFT schemes in [30] and [31] rely heavily on dispersion, in which the frequency resolution is limited by the dispersion value. To reduce the dispersion requirement for high-frequency resolution, a bandwidth-scaled-microwave-photonics-based STFT scheme, consisting of electronic-to-optical conversion, temporal sampling, and wavelength-to-time mapping (WTM), was proposed in [32], with a frequency resolution of 60 MHz, an observation bandwidth of 1.98 GHz, and an acquisition frame rate of 160 MHz. Thanks to the use of bandwidth-scaled electrical-to-optical conversion with the help of a multi-wavelength source and a periodic optical filter, the dispersion requirement, as well as the transmission delay, is greatly reduced. However, a temporal sampling unit is needed to enable the time-domain separation of different wavelengths in the WTM process. Furthermore, the time lens is needed to avoid broadening of the pulses after propagation in the DCF. All in all, the reported photonic-assisted STFT schemes are all based on dispersion devices, such as the LCFBG and the DCF, which suffer from the performance limitation imposed by the dispersion medium used. On the one hand, the frequency resolution of the dispersion-based all-optical STFT schemes in [30], [31] is very demanding for dispersion; on the other hand, the dispersion devices are hard to be reconstructed in real-time in multiple dimensions, which severely constrains the reconfigurability, though the scheme in [32] reduces the need of dispersion values for implementing a high-frequency resolution. In addition, the analysis bandwidths in [30] and [31] are limited to no more than 2.5 GHz, which needs to be further improved.

In optical fibers, one of the nonlinear effects nominated stimulated Brillouin scattering (SBS) caused by the acousto-optic interaction, is of great concern [33]-[42] due to the Brillouin gain provided by it.

The SBS gain owns a typical narrow bandwidth of about dozens of MHz with good tunability and can be utilized to selectively amplify the optical signals and enable a microwave photonic filter. Various SBS-based microwave frequency measurement proposals have also been widely investigated [39]-[42]. Recently, by introducing a two-step accuracy improvement, we proposed an accuracy-improved multiple-frequency microwave measurement approach using SBS-based frequency-to-time mapping (FTTM) [43], with the measurement accuracy greatly improved to better than ±1 MHz. To further improve the 20-MHz frequency resolution in [43], we proposed another multiple microwave frequency measurement approach using a narrowed SBS gain [44], with a measurement resolution of better than 10 MHz. However, up to now, we note that few SBS-based frequency measurement approaches are designed for broadband microwave signals, let alone time-frequency analysis of broadband signals. In 2017, an SBS-based approach for the frequency measurement of broadband signals was proposed [45]. However, only periodic microwave signals can be measured using this approach and the measurement is not in real-time. Based on our previous SBS-based microwave frequency measurement proposals [43], [44], we make efforts to implement the real-time all-optical STFT for the time-frequency analysis of broadband microwave signals using the SBS effect.

In this paper, we propose and experimentally demonstrate an all-optical STFT scheme based on SBS-based FTTM. The work presented is primarily motivated by the need to avoid the limitation of the reported dispersion-based STFTs. In the proposed all-optical STFT scheme, a key and ingenious design is that the SBS not only provides a band-pass filter for implementing the window function in conjunction with a periodic frequency-sweep optical signal but also obtains the frequency domain information in different time windows through the generated waveform via the SBS-based FTTM. A periodic frequency-sweep optical signal is generated and then modulated by an electrical SUT at a Mach-Zehnder modulator (MZM). During different sweep periods, the fixed Brillouin gain functions as a bandpass filter to filter a specific range of the spectrum, which is equivalent to applying a sliding window function to the corresponding section of the temporal signal with the help of the sweep optical signal. At the same time, after the optical signal is selectively amplified by the SBS gain and converted back to the electrical domain, SBS also implements the real-time FTTM, which can be utilized to obtain the frequency domain information corresponding to different time windows through the generated waveforms via the FTTM. The frequency domain information corresponding to different time windows is obtained and spliced to analyze the time-frequency relationship of the SUT. An experiment is performed. STFTs of a variety of RF signals are carried out, and the dynamic frequency resolution is 60 MHz. The 12-GHz observation bandwidth in the experiment is limited only by the adopted equipment. The real-time performance of the proposed STFT scheme is also demonstrated by capturing a burst signal with a duration of 2 μs.

## 2. Principle
### 2.1. Theoretical analysis

Fig. 1(a) graphically illustrates the operation of STFT. STFT examines the frequency components as a window moves in time, which truncates the whole time-domain signal into a series of shorter-duration signals until the window reaches the end of the signal. The time duration should be short enough to enable that the signal inside the window can be considered stationary. Then, each windowed signal is transformed to the frequency domain by computing the Fourier transform (FT). Each FT provides the spectral information of a separate time-slice of the signal, providing simultaneous time and frequency information. Fig. 1(b) shows a block diagram for the implementation of STFT, which demonstrates the process of STFT for the input SUT. A finite-length temporal window function $h(t)$ is chosen and multiplied by the SUT $s(t)$ to keep a certain portion of it. Then, FT is adopted to the windowed SUT and the results are saved by calculating integrals. Mathematically, STFT is defined as

$$STFT_s(\tau,\omega) = FT[h(t-\tau)s(t)]$$
$$= \int_{-\infty}^{+\infty} h(t-\tau)s(t)\exp(-j\omega t)dt, \quad (1)$$

where $h(t)$ is a window function for calculation of STFT, $\omega$ represents the angular frequency variable, and $\tau$ is a time-delay parameter that defines the STFT analysis time axis and typically extends over the entire duration of the SUT $s(t)$. Fig. 1(c) briefly shows the schematic of the proposed all-optical real-time STFT approach, which involves a cascade of an optical domain multiplier unit implemented by an MZM to modulate the SUT envelope by a periodic sweep optical signal, and a frequency-dependent SBS-based FTTM unit to obtain the frequency domain information during each sweep period. The temporal windowing is realized during the periodic sweep process. The entire process includes two windowing processes: the first windowing is to perform windowing in the time domain by using the sweep optical signal; the second windowing is to use SBS to perform windowing in the frequency domain for each windowed signal during each sweep period. In each short sweep period, after the optical signal is selectively amplified by the SBS gain and converted back to the electrical domain, the spectrum information can be obtained in real-time through the SBS-based FTTM. Through periodic scanning, the spectrum information of different scanning periods can be obtained. In fact, STFT has been completed during the periodic sweep process with the help of SBS-based FTTM, where the sliding temporal window function is implemented by using the fixed Brillouin gain and the sweep process. SBS not only provides a band-pass filter for forming window function in the second windowing process but also extracts the frequency domain information in different time windows.

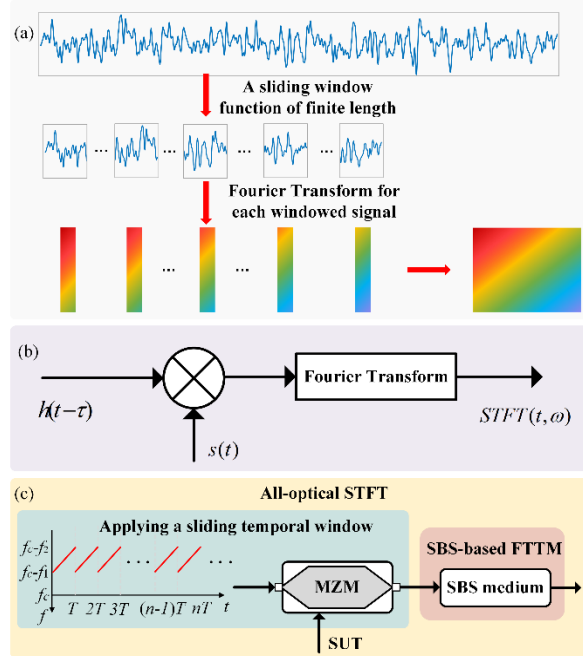

Fig. 1. (a) Illustration of the concept of STFT, (b) block diagram of the realization of STFT, (c) the schematic of the all-optical STFT approach, consisting of an optical domain multiplier unit (MZM) to modulate the SUT envelope by a periodic sweep optical signal, and a frequency-dependent SBS-based FTTM to obtain the frequency domain information during each sweep period (each sweep period is regarded as a temporal window).

It is assumed that the sweep period $T$ is small enough so that local stationarity for the SUT after being windowed is achieved during each sweep period. The SUT $s(t)$ is modulated on the sweep optical signal at the MZM, where the RF spectrum is converted to the optical domain. Here, the sweep optical signal in a period can be obtained by carrier-suppressed lower single-sideband (CS-LSSB) modulation and expressed as

$$s_A(t) \propto \exp[j\omega_c t - j(\omega_0 + \frac{1}{2}kt)t] \\ = 1 \times \exp[j(\omega_0' - \frac{1}{2}kt)t], t \in (0, T], \quad (2)$$

where $\omega_c$ is the angular frequency of the optical carrier, $\omega_0$ is the original angular frequency of the electrical sweep signal, $\omega_0' = \omega_c - \omega_0$, $k = 2\pi K$, and $K$ is the sweep chirp rate of the electrical sweep signal.

At the same time, a temporal window $h(t)$, centered at zero, is adopted and its first period can be expressed as

$$h(t) = \begin{cases} 1, t \in (-\frac{T}{2}, \frac{T}{2}] \\ 0, otherwise \end{cases}. \quad (3)$$

In combination with Eq. (3), Eq. (2) is rewritten as

$$s_A(t) \propto h(t - \frac{T}{2}) \exp[j(\omega_0' - \frac{1}{2}kt)t], \quad (4)$$

which is multiplied by the SUT to add a time window to the SUT. Then, the SUT after adding the time-domain window function can be expressed as

$$s_B(t) \propto s_A(t) s(t) \\ = h(t - \frac{T}{2}) s(t) \exp[j(\omega_0' - \frac{1}{2}kt)t], \quad (5)$$

We might as well define a new function $a(t)$ expressed as

$$a(t) = h(t - \frac{T}{2}) s(t). \quad (6)$$

The spectrum of $a(t)$ can be expressed as

$$A(\omega) = FT[a(t)] = \int_{-\infty}^{+\infty} h(t - \frac{T}{2}) s(t) \exp(-j\omega t) dt \\ = STFT_s(\frac{T}{2}, \omega). \quad (7)$$

Thus, the frequency spectrum of $s_B(t)$ can be expressed as

$$S_B(\omega) \propto \int_{-\infty}^{+\infty} h(t - \frac{T}{2}) s(t) \exp[j(\omega_0' - \frac{1}{2}kt)t] \exp(-j\omega t) dt \\ = FT[a(t) \exp[j(\omega_0' - \frac{1}{2}kt)t]]. \quad (8)$$

Here, as mentioned above, as the sweep period $T$ is small enough, local stationarity is achieved. That is, the frequency in the time window can be considered as a fixed value and expressed as

$$\omega^\dagger = \omega_0' - kt, \quad (9)$$

Thus, Eq. (8) can be rewritten as

$$S_B(\omega) \propto FT[a(t)\exp(j\omega^\dagger t)] = A(\omega - \omega_0' + kt). \tag{10}$$

Then, SBS gain selectively extracts the specific spectral information during each sweep period, which can be considered as adding a window function in the frequency domain. During each sweep process, the spectrum output from each sweep period can be expressed as

$$Y(\omega) \propto A(\omega - \omega_0' + kt)g_B(\omega), \tag{11}$$

where $g_B(\omega)$ represents the Lorentzian Brillouin gain profile created by a continuous pump wave and can be expressed as [46]

$$g_B(\omega) = g_{B0}\frac{(\Delta\Omega_B/2)^2}{(\Delta\Omega_B/2)^2 + (\omega - \omega_B)^2}, \tag{12}$$

where the gain profile is around the red-shifted frequency $\omega_B = \omega_p - \Omega_B$, $\Omega_B$ is the Brillouin shift in the fiber, $\omega_p$ is the frequency of the pump wave, $\Delta\Omega_B$ is the full width at half-maximum (FWHM) of the gain profile, and $g_{B0}$ is the peak value of the Brillouin gain.

The temporal expression of the output corresponding to the first sweep period can be obtained by inverse Fourier transform of $Y(\omega)$, which can be expressed as

$$\begin{aligned} y(t) &= IFT[Y(\omega)] \\ &\propto \int_{-\infty}^{+\infty} g_B(\omega)A(\omega - \omega_0' + kt)\exp(j\omega t)d\omega. \end{aligned} \tag{13}$$

For simplicity, the Brillouin gain is considered as the Dirac function to demonstrate the SBS-based FTTM process. Thus, Eq. (13) can be rewritten as

$$\begin{aligned} y(t) &\approx \int_{-\infty}^{+\infty} \delta(\omega - \omega_B)A(\omega - \omega_0' + kt)\exp(j\omega t)d\omega \\ &= A(\omega_B - \omega_0' + kt)\exp(j\omega_B t). \end{aligned} \tag{14}$$

As can be seen from Eq. (14), the spectrum of the first sweep period is mapped to the time domain. Different time from 0 to T corresponds to different angular frequencies from $\omega_B - \omega_0'$ to $\omega_B - \omega_0' + kT$.

When the optical signal in Eq. (14) is detected by a photodetector (PD), the photocurrent can be expressed as

$$\begin{aligned} i(t) \propto |y(t)|^2 &= \left|A(\omega_B - \omega_0' + kt)\right|^2 \\ &= \left|STFT(\frac{T}{2}, \omega_B - \omega_0' + kt)\right|^2. \end{aligned} \tag{15}$$

From Eq. (15), the amplitude information of STFT of the first sweep period can be obtained. STFTs of different sweep periods are then separated in the time domain by FTTM and further processing. By collecting the spectrum information of each scanning period in the time domain and stacking it into a two-dimensional time-frequency plane in chronological order, STFT of the SUT can be obtained.

*2.2. Photonic implementation of STFT*

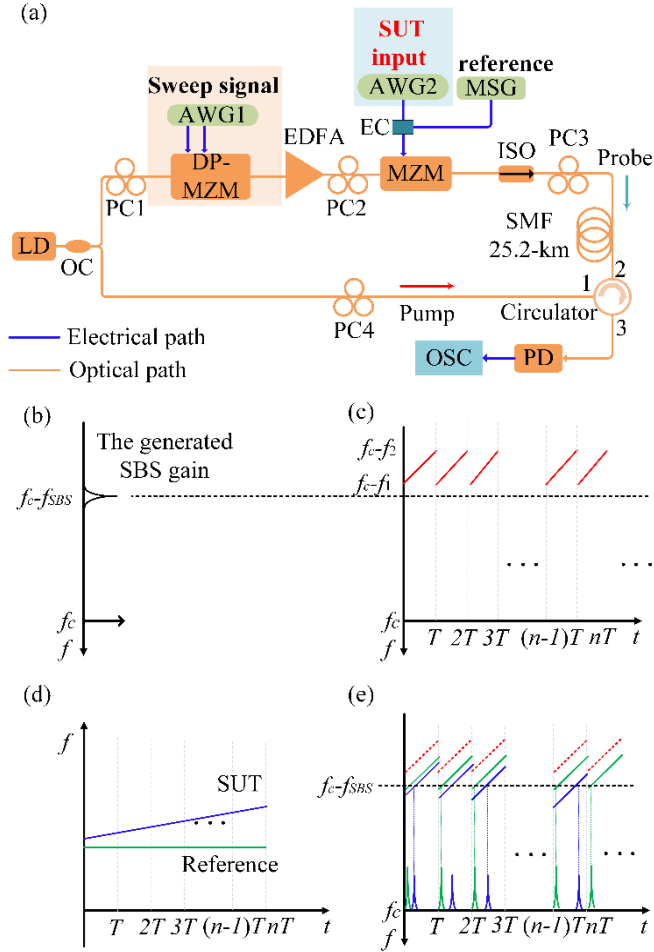

Fig. 2. (a) Schematic diagram of the proposed SBS-based real-time all-optical STFT approach. LD, laser diode; OC, optical coupler; PC, polarization controller; DP-MZM, dual-parallel Mach-Zehnder modulator; MZM, Mach-Zehnder modulator; AWG, arbitrary waveform generator; MSG, microwave signal generator; EDFA, erbium-doped fiber amplifier; ISO, isolator; SMF, single-mode fiber; PD, photodetector; OSC, oscilloscope. (b) The generated SBS gain with its frequency centered at $f_c$-$f_{SBS}$ by the optical carrier centered at $f_c$. (c) Time-frequency characteristics of the generated sweep optical signal from the DP-MZM, where $T$ is the sweep period. (d) Time-frequency characteristics of the SUT and a single-tone reference. (e) Time-frequency characteristics and the FTTM of the optical signal from the optical circulator.

Fig. 2(a) shows the schematic diagram of the proposed all-optical real-time STFT. A continuous-wave (CW) light wave from a laser diode (LD) is split into two branches. In the lower branch, the CW light wave as the pump wave is injected into a spool of single-mode fiber (SMF) via an optical circulator, which will generate an SBS gain with its frequency centered at $f_c$-$f_{SBS}$ as shown in Fig. 2(b). In the upper branch, the CW light wave is CS-LSSB modulated at a dual-parallel Mach-Zehnder modulator (DP-MZM) by an electrical sweep signal (from $f_1$ to $f_2$) to generate a periodic sweep optical carrier, with its time-frequency characteristic shown in Fig. 2(c), which has a period of $T$, a positive chirp rate of $k$, and start frequency and stop frequency of $f_c$-$f_1$ and $f_c$-$f_2$, respectively. Subsequently, the sweep optical carrier is modulated at a null-biased MZM by a combined signal, which includes an SUT and a single-frequency reference at $f_r$ with the time-frequency characteristic shown in Fig. 2(d). Ideally, considering the idea of limit in mathematics, the SUT can be divided infinitely in time, and the frequency in each short time can be regarded as a specific value. Then, the optical signals from the MZM are sent to the SMF via the isolator (ISO) as the probe wave, which will detect the SBS gain.

Because the MZM is null-biased, the output of the MZM is a CS-DSB signal. As shown in Fig. 2(e), the negative sidebands that do not interact with the SBS gain during the scanning process are not shown. The reference frequency $f_r$ and the start frequency $f_1$ are designed to satisfy the following condition

$$f_1 - f_r = c, \tag{16}$$

where $c$ is a constant and just slightly smaller than $f_{SBS}$. In this case, the positive sidebands of the single-frequency reference can be just amplified by the SBS gain at a certain time in every period during the scanning process. Pulses in different periods (time windows) represent the frequency components in the corresponding time windows. Then, the pulses generated in the time domain through the FTTM are recombined to obtain the time-frequency diagram of the SUT. Here, a known reference signal is applied to the system along with the SUT, which is only used to provide a label of time and frequency during the recombination.

## 3. Experimental results and discussion
### 3.1. Experimental setup

A proof-of-concept experiment based on the setup shown in Fig. 2(a) is performed to verify the proposed all-optical real-time STFT scheme. A 15.5-dBm optical carrier centered at 1551.303 nm from the LD (ID Photonics, CoBriteDX1-1-C-H01-FA) is divided into two paths via a 50/50 optical coupler (OC). In the upper path, the optical carrier is CS-LSSB modulated at the DP-MZM (Fujitsu, FTM 7961EX) by a sweep electrical signal from AWG1 (Keysight M8195A) to generate a sweep optical signal. The output of the DP-MZM is injected into the null-biased MZM (Fujitsu, FTM 7938) after being amplified by the EDFA (Amonics, EDFA-PA-35-B), and then CS-DSB modulated by a combined signal from the electrical coupler (EC, Narda 4456-2), which includes the SUT from AWG2 (Keysight M8190A) and a fixed reference from the microwave signal generator (MSG, Agilent 83630B). Then, the output of the MZM is injected into the 25.2-km SMF through the ISO. Polarization controllers (PC1 and PC2) are used to optimize the light polarizations before the DP-MZM and MZM, respectively. In the lower path, the CW light wave from the LD is used as the pump wave and launched into the 25.2-km SMF via an optical circulator, where it interacts with the counter-propagating probe wave from the upper branch. PC3 and PC4 are used to ensure the most efficient stimulated Brillouin interaction. Then, the optical signal from the SMF is detected by a PD (Nortel PP-10G) and monitored by an oscilloscope (OSC, Rohde & Schwarz, RTO2032).

### 3.2. Analysis bandwidth and frequency range

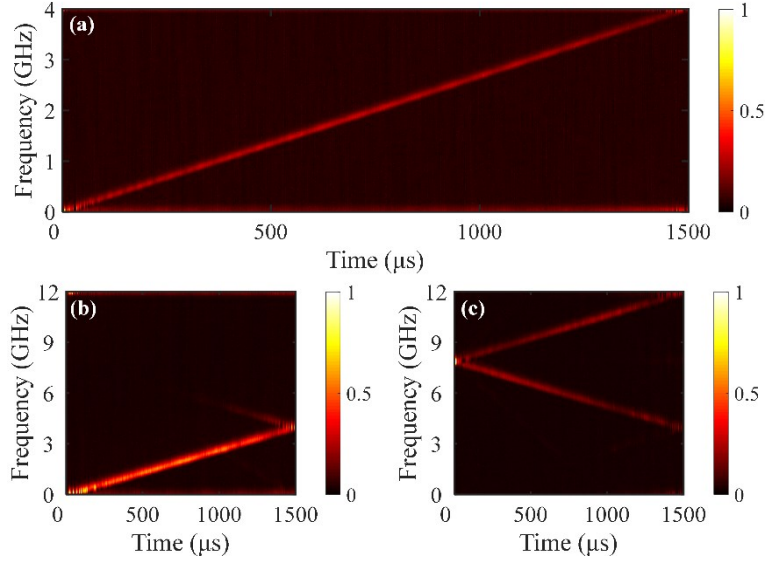

Fig. 3. Measured time-frequency diagrams of the LFM signal with a time duration of 1500 μs and a bandwidth ranging from 10 MHz to 4 GHz by the proposed STFT scheme configured with an electrical sweep signal from (a) 10.8 to 14.8 GHz, (b) 10.8 to 22.8 GHz. (c) Measured time-frequency diagram of the LFM signal after it is mixed with an LO signal at 8 GHz by using an electrical sweep signal from 10.8 to 22.8 GHz.

The analysis bandwidth and frequency range offered by the proposed STFT scheme are demonstrated. The period, bandwidth, and center frequency of the electrical sweep signal from AWG1 are set to 2 μs, 4 GHz, and 12.8 GHz, respectively. The 0-dBm CW reference signal from the MSG is fixed at 10 MHz unless otherwise specified in this paper. It is expected that the analysis frequency range under such parameter settings is from 0 to 4 GHz, taking into account the stimulated Brillouin frequency shift $f_B$ of around 10.8 GHz in the experiment. Accordingly, a linearly frequency-modulated (LFM) signal with a bandwidth ranging from 10 MHz to 4 GHz and a time duration of 1500 μs from AWG2 is chosen as the SUT, and the measured time-frequency diagram by the proposed STFT scheme is shown in Fig. 3(a), which is in line with the expectation. To further illustrate the large bandwidth analysis capabilities, the scheme is configured with another electrical sweep signal, with the period, bandwidth, and center frequency of 2 μs, 12 GHz, and 16.8 GHz, respectively. In this case, the analysis frequency range is from 0 to 12 GHz. The same SUT with a time duration of 1500 μs and a bandwidth ranging from 10 MHz to 4 GHz is chosen, and the measured time-frequency diagram is shown in Fig. 3(b), which is consistent with the result in Fig. 3(a). Due to AWG2 limitations, the highest frequency that can be generated is limited to around 4 GHz. To verify the STFT in the frequency range from 0 to 12 GHz, the LFM signal from 10 MHz to 4 GHz is mixed with an 8-GHz LO signal, resulting in the generation of a dual-chirp LFM signal with a positive chirp from 8.01 to 12 GHz and a negative chirp from 7.99 to 4 GHz. Fig. 3(c) shows the measured time-frequency diagram of the dual-chirp LFM signal, which indicates that a signal up to 12 GHz can be well analyzed.

From the results presented in Fig. 3, the analysis bandwidth is determined by the bandwidth of the electrical sweep signal and is only limited by the adopted equipment. Another issue worth mentioning is that the analysis bandwidth and frequency range of the proposed STFT scheme can be easily reconfigured. In our proposal, the analysis frequency range is from $f_B$-$f_1$ to $f_B$-$f_1$+$KT$ based on Eqs. (14)-(15), which can be reconfigured by changing the frequency range of the electrical sweep signal. In addition, changing the pump wavelength can also reconfigure the analysis frequency range. In this work, the pump wave is directly tapped from the LD. To shift the analysis frequency range, a frequency shift module can be added in the lower path from the OC, or an individual pump laser with its wavelength properly set is used as the pump wave. Because the tunable analysis bandwidth and frequency range is demonstrated in Fig. 3, in the following part of the work, the scheme is further demonstrated and analyzed in the frequency range from 0 to 4 GHz.

*3.3. Analysis resolution*

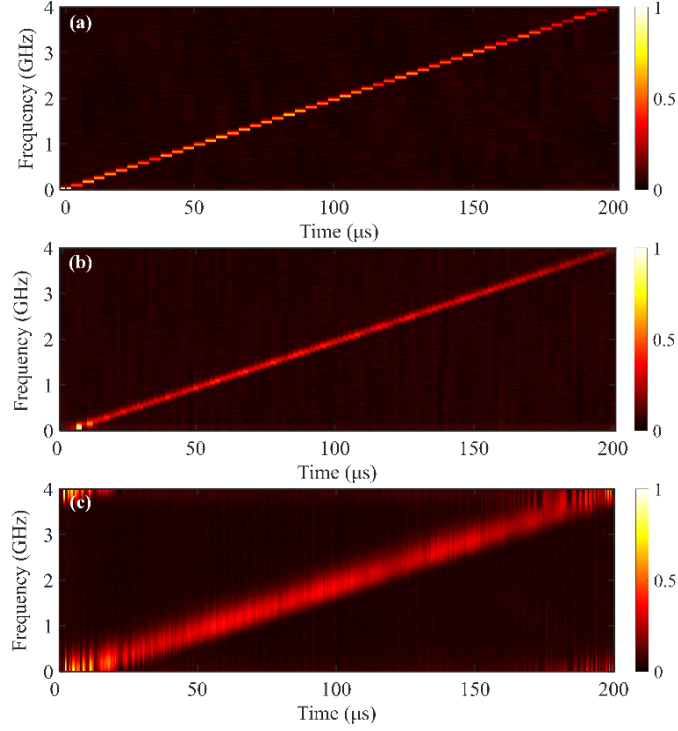

Fig. 4. Measured time-frequency diagrams of the LFM signal with a bandwidth ranging from 10 MHz to 4 GHz and a time duration of 200 μs by the proposed STFT scheme configured with an electrical sweep signal with (a) 4-μs period, (b) 2-μs period, (c) 0.5-μs period.

The analysis resolution both in frequency and time are then studied. First, the resolution of the proposed STFT scheme configured with different parameters (i.e. the sweep chirp rate or sweep period) to obtain the time-frequency information of the same SUT is demonstrated. To this end, the scheme is configured with three different electrical sweep signals with the same frequency ranging from 10.8 to 14.8 GHz and different sweep periods of 4, 2, and 0.5 μs, so the sweep chirp rates are 1, 2, and 8 GHz/μs, respectively. An LFM signal with a bandwidth ranging from 10 MHz to 4 GHz and a time duration of 200 μs is chosen as the SUT. The measured time-frequency diagrams are shown in Fig. 4. As can be seen, when the sweep bandwidth is fixed, with the decrease of the sweep period or the increase of the chirp rate, the time resolution will increase, while the frequency resolution will decline. The reason is that the frequency resolution is related to the width of the time-domain pulses obtained through the SBS-based FTTM, which will be broadened as the sweep chirp rate increases and the corresponding SBS interaction time decreases at a certain frequency point. Said other words, the time-domain pulses will overlap each other and cannot be distinguishable as the chirp rate is too large. Thus, a smaller sweep chirp rate $K$ is needed to enable the time-domain separation of different frequencies in the FTTM process to guarantee a good frequency resolution of STFT. However, for a certain analysis bandwidth, a smaller sweep chirp rate means a longer period, which may destroy the short-term signal stationarity of the SUT in a sweep period and in turn reduce the measurement accuracy. It is concluded that a tradeoff between the time resolution and the frequency resolution is present in the proposed STFT scheme, which is consistent with the characteristic of STFT.

In addition, it is noticed that the analysis results in terms of time resolution and frequency resolution for the same LFM signal are quite different. Fig. 4(a) even shows a relatively visible error, which indicates that it is necessary to select a sweep signal with an appropriate chirp rate to obtain the desired result according to different application scenarios and requirements. From the measured results presented in Fig. 4, another issue that should be mentioned is that the resolution performance is unchanged for analyzing the same SUT, as the window length is fixed. Fortunately, the resolution performance can be easily reconstructed by simply changing the period of the sweep optical signal. In practice, the SUT can be firstly measured by the proposed scheme configured with a sweep electrical signal with relatively large bandwidth and sweep chirp rate. After the SUT is measured coarsely, the proposed scheme can be configured with a sweep electrical signal with an appropriate chirp rate to achieve a relatively accurate analysis if it is needed.

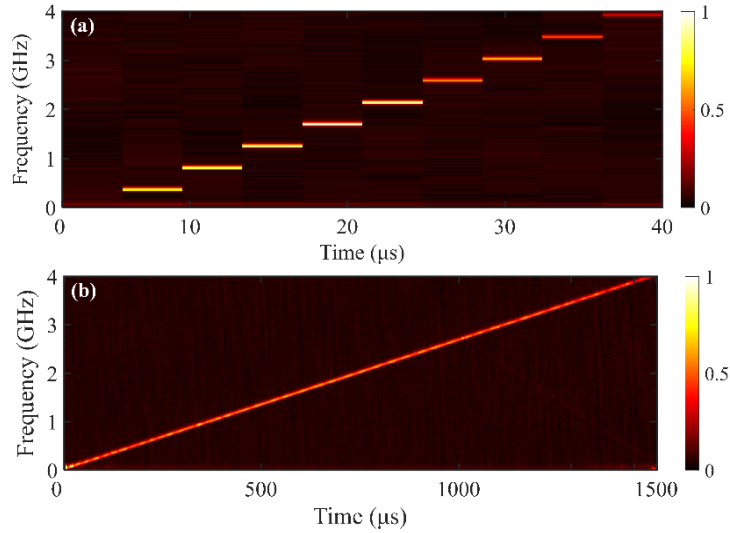

Fig. 5. Measured time-frequency diagrams of the LFM signals with a bandwidth ranging from 10 MHz to 4 GHz and a time duration of (a) 40 μs and (b) 1500 μs by the proposed STFT scheme configured with a certain electrical sweep signal with a fixed sweep chirp rate of 1 GHz/μs, a period of 4 μs, and a bandwidth ranging from 10.8 to 14.8 GHz.

The resolution performance of the proposed STFT scheme configured with the same parameters when acquiring time-frequency information of different SUTs is studied. To this end, the scheme is configured with a certain electrical sweep signal with a fixed sweep chirp rate of 1 GHz/μs and a bandwidth ranging from 10.8 to 14.8 GHz. LFM signals with a bandwidth ranging from 10 MHz to 4 GHz and time durations of 40 and 1500 μs are chosen as the SUTs, respectively, with the measured results shown in Fig. 5. Compared the measured result of Fig. 5 with that of Fig. 4(a), one can easily find that the resolution performance of the proposed STFT scheme is also highly related to SUT under the specific parameter configuration of the scheme. When a SUT with a shorter time length is analyzed, the period of the sweep signal also needs to be shortened to guarantee an acceptable time-frequency diagram. However, as shown in Fig. 4, the period of the sweep signal cannot be arbitrarily decreased due to that a certain interaction time with the pump wave is needed to excite the SBS effect to generate a distinguishable pulse.

Then, the frequency resolution of the scheme is verified using four two-tone signals. To do this, the scheme is configured with different sweep signals having a sweep chirp rate of 1, 1.333, 1.6, and 2 GHz/μs, respectively. The measured time-frequency diagrams of the two-tone signals are shown in Fig. 6, where the frequency resolutions are better than 60, 85, 100, and 116 MHz, respectively. As can be seen, the frequency resolution becomes worse as the sweep chirp rate increases. The reason is that the time-domain pulses obtained through SBS-based FTTM have a certain width due to the FWHM of the SBS gain profile, which will be broadened as the sweep period decreases, resulting in a poorer frequency resolution. Furthermore, even if the sweep chirp rate is very small, the resolution cannot be arbitrarily increased because the SBS gain has a certain bandwidth, which limits the lower limit of the width of the generated pulses and also the upper limit of the frequency resolution.

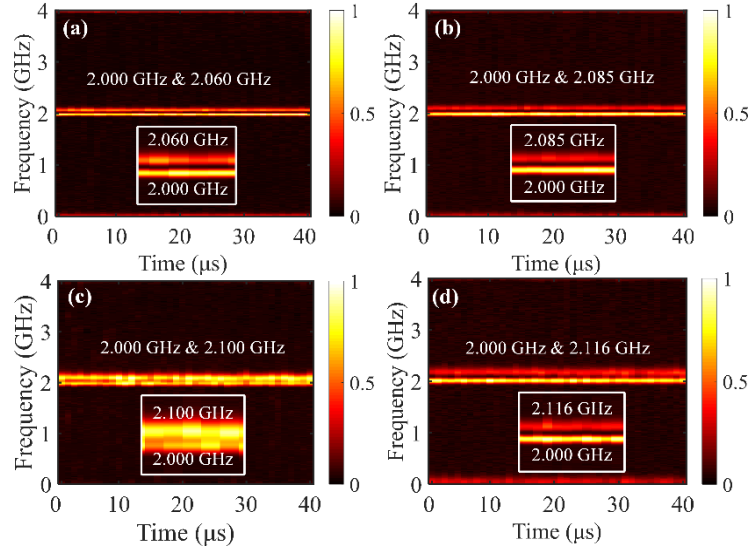

Fig. 6. Measured time-frequency diagrams by the proposed STFT scheme for the two-tone signal with frequencies of (a) 2 and 2.060 GHz, (b) 2 and 2.085 GHz, (c) 2 and 2.100 GHz, and (d) 2 and 2.116 GHz.

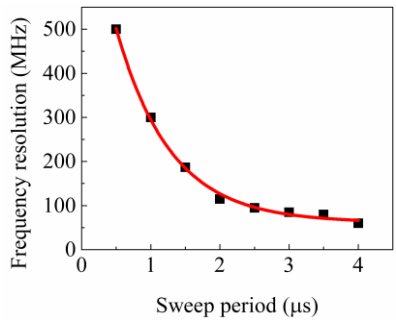

Fig. 7. Frequency resolution varied with the sweep period when the sweep bandwidth is fixed at 4 GHz. The red solid line is a fitted curve.

To better confirm the influence of sweep signals on frequency resolution, more experiments are carried out. Fig. 7 shows that the frequency resolution is improved with the increase of the sweep period when the sweep bandwidth is fixed at 4 GHz. To a certain extent, the longer the sweep period, the better the frequency resolution, but when the sweep period approaches a certain value, the resolution approaches a limit. However, for a certain sweep bandwidth, it will lead to poorer time resolution and even cause an obvious error due to not satisfying the local stationarity. It should be mentioned that narrowing the SBS gain bandwidth can further improve the frequency resolution [44].

*3.4. Analysis of multi-format signals*

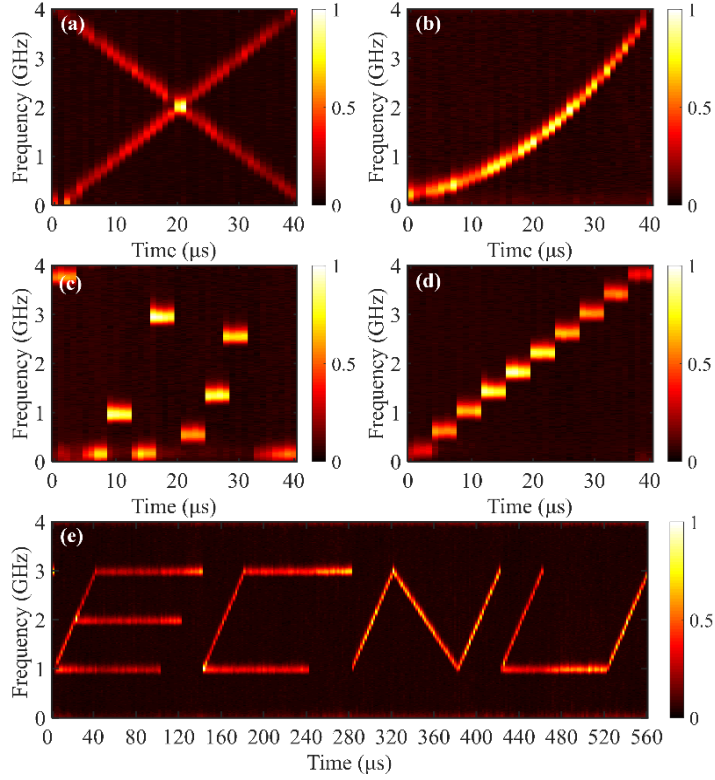

Fig. 8. Measured time-frequency diagrams by the proposed STFT scheme for (a) dual-chirp LFM signal, (b) non-linearly frequency modulated signal, (c) frequency-hopping signal, (d) step-frequency signal, (e) the time-varied signal whose time-frequency diagram is the word "ECNU".

To further verify the time-frequency analysis capability of the proposed STFT scheme, time-frequency analysis of multi-format signals is further demonstrated. The period, bandwidth, and center frequency of the electrical sweep signal from AWG1 are set to 1 μs, 4 GHz, and 12.8 GHz, respectively. Different kinds of broadband signals, including dual-chirp LFM signal, non-linearly frequency modulated (NLFM) signal, frequency-hopping (FH) signal, and step-frequency (SF) signal, are chosen as the SUTs. The time duration of each type of signal is 40 μs and the signal bandwidth covers the frequency range from 0 to 4 GHz. The results of the time-frequency analysis for those four signals are shown in Fig. 8 (a)-(d). Fig. 8(e) shows the measured time-frequency diagram for a 560-μs signal whose time-frequency relationship is designed as the abbreviation "ECNU" of East China Normal University. The proposed STFT scheme is configured with an electrical sweep signal, with the period, bandwidth, and center frequency of 2 μs, 4 GHz, and 12.8 GHz, respectively. As can be seen, the time-frequency diagrams of all these kinds of signals are well constructed with good resolution, which confirms the capability of the STFT scheme for the time-frequency analysis of multi-format signals.

*3.5. Burst detection*

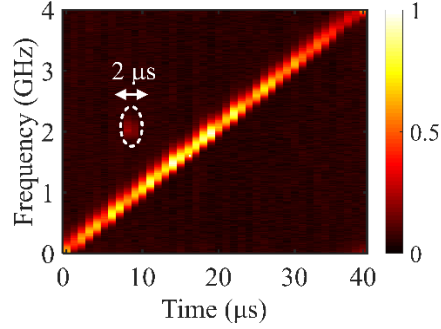

Fig. 9. Measured time-frequency diagram of the SUT consisting of an LFM from 10 MHz to 4 GHz and a burst signal with a duration of 2 μs and a carrier frequency of 2 GHz.

To further illustrate the real-time performance of the proposal, the ability to detect a burst signal is demonstrated. The period, bandwidth, and center frequency of the sweep signal from AWG1 are set to 1 μs, 4 GHz, and 12.8 GHz, respectively. The input RF signal, consisting of an LFM from 10 MHz to 4 GHz and a burst signal with a duration of 2 μs and a carrier frequency of 2 GHz, is chosen as the SUT. As shown in Fig. 9, the burst signal from 8-10 μs is well detected. The burst in the time-frequency diagram is weak because the power of the burst signal is 8 dB lower than that of the LFM signal. From the results in Fig. 9, the burst detection capability is well verified, which further confirms the good real-time performance of the scheme. However, due to the use of a fixed window length and a fixed frequency sweep chirp rate in the fixed window, the burst that can be detected needs to meet the condition that the time length of the burst is greater than the sweep period to ensure that it can be captured by the SBS gain.

*3.6. Comparison*

A comparison between the proposed real-time all-optical STFT scheme and the previously reported ones [30]-[32] is shown in Table I. The reported STFTs are all based on dispersion devices, such as the LCFBG and the DCF. The STFT schemes in [30], [31] need large dispersion values to obtain a high frequency resolution, so they suffer from performance limitations mainly imposed by the relatively limited amount of dispersion offered by the dispersion medium. As can be seen from Table 1, the analysis bandwidth in [30]-[32] is no more than 2.43 GHz. In comparison, in this work, the analysis bandwidth is demonstrated to be 12 GHz, which can be further increased by using a larger-bandwidth electrical

Table 1. Comparison of different photonics-assisted time-frequency analysis methods

| Proposal | STFT | Techniques/ Devices | Dispersion dependent | Complexity | Bandwidth | Frequency Resolution | Time Resolution | Real-time Availability | Reconfigur-ability |
|---|---|---|---|---|---|---|---|---|---|
| Ref. [30] | Yes | LCFBG Array | Yes | Middle | Unspecified | | | Yes | Low |
| Ref. [31] | | Time sampling & DCF | | Low | 2.43 GHz | 340 MHz | 5.9 ns | | Middle |
| Ref. [32] | | Time lens & DCF | | High | 1.98 GHz | 60 MHz | 6.25 ns | | |
| Ref. [45] | No | SBS-FTTM | No | Middle | 27.5 GHz | Unspecified | | No | High |
| This work | Yes | | | Middle | 12 GHz | 60 MHz* | 0.5 μs* | Yes | |

* The best results respectively demonstrated in this work. The frequency resolution and time resolution have a tradeoff as the DSP-based STFT does.

sweep signal. Moreover, the dispersion devices are hard to be reconstructed in real-time in multiple dimensions, which severely constrains the reconfigurability of these approaches in [30]-[32]. Although an approach for the time-frequency analysis is proposed based on SBS [45], it can only be used to measure the periodic signal, which is not in real-time. In this work, STFT is in real-time. Although a 25.2-km SMF functions as the nonlinear medium for the SBS effect, it can be replaced by a much shorter highly nonlinear fiber, which guarantees it good real-time performance as the scheme in [32]. Thanks to the narrow and flexible gain spectrum enabled by the SBS effect and a simpler SBS-pump and probe scheme with the help of a sweep optical signal, the proposed STFT scheme not only can be reconstructed in multiple dimensions (bandwidth, frequency range, frequency resolution, and time resolution) by simply changing the electrical sweep signal, the reference and pump frequency, but also provides a significant performance increase in the operating frequency, bandwidth over the reported ones [30]-[32]. As discussed in Eq. (15), the proposed scheme is theoretically an STFT scheme, so there is a tradeoff between the frequency resolution and the time resolution. Furthermore, the time resolution of this work is worse than that in [31], [32]. We believe that the scheme proposed in this paper is more suitable for the time-frequency analysis of signals with large bandwidth, long term, and relatively slow changes in the case where very high reconfigurability is required.

## 4. Conclusions

In summary, we have proposed and experimentally demonstrated an all-optical STFT scheme based on SBS-based FTTM. The key significance of the work is that STFT is implemented by using the SBS effect for the first time and further verified by incorporating it in the time-frequency analysis of multi-format microwave signals, to the best of our knowledge. Thanks to the SBS-based FTTM, which not only provides a band-pass filter for implementing the window function of STFT in conjunction with a periodic frequency-sweep optical signal but also obtains the frequency domain information in different time windows through the generated waveform via the SBS-based FTTM. The proposed all-optical real-time STFT scheme overcomes the problem that the existing photonics-assisted STFT schemes based on dispersive medium need a large dispersion value and have poor reconfigurability and limited analysis bandwidth. A proof-of-concept experiment is performed to verify its feasibility. STFTs of a variety of RF signals are carried out, and the dynamic frequency resolution is 60 MHz and the 12-GHz analysis bandwidth is limited only by the equipment.

**Funding**

Natural Science Foundation of Shanghai (20ZR1416100); National Natural Science Foundation of China (NSFC) (61971193); Open Fund of State Key Laboratory of Advanced Optical Communication Systems and Networks, Peking University, China (2020GZKF005); Science and Technology Commission of Shanghai Municipality (18DZ2270800).

**Conflicts of interest**

The authors declare no conflicts of interest.